\documentclass[twocolumn,showpacs,preprintnumbers,amsmath,amssymb,prl]{revtex4-1}
\pdfoutput=1

\usepackage{graphicx}
\usepackage{dcolumn}
\usepackage{bm}
\usepackage{natbib} 
\usepackage{textcomp}

\begin{document}

\preprint{APS/123-QED}

\title{Fully electrically read-write device out of a ferromagnetic semiconductor}

\author{S. Mark, P. D\"urrenfeld, K. Pappert, L. Ebel, K. Brunner, C. Gould, and L.W. Molenkamp}
 \affiliation{Physikalisches Institut (EP3) and R\"ontgen Center for Complex Material Systems, Am Hubland, Universit\"at W\"urzburg, D-97074 W\"urzburg, Germany}
\date{\today}

\begin{abstract}

We report the realization of a read-write device out of the ferromagnetic semiconductor (Ga,Mn)As as the first step to fundamentally new information processing paradigm. Writing the magnetic state is achieved by current-induced switching and read-out of the state is done by the means of the tunneling anisotropic magneto resistance (TAMR) effect. This one bit demonstrator device can be used to design a electrically programmable memory and logic device.

\end{abstract}


\maketitle

At present memory and logic fabrication are two fully separated architectures \cite{Awschalom2007,Wolf2001}. While bulk information storage traditionally builds on metallic ferromagnets, logic makes use of gateability of charge carriers in semiconductors. Combining storage and processing in a single monolithic device not only would solve current technical issues such as the heat dissipation generated by transferring information between the two architectures, but also offer the possibility of a fully non-volatile information processing system. Here we present a read-write device which can be used as one element of an electrically programmable logic gate. Our structure is made from the ferromagnetic semiconductor (Ga,Mn)As, which exhibits carrier-induced ferromagnetism at low temperatures \cite{Dietl2000,Abolfath2001,Ohno1998}. The 70 nm thick (Ga,Mn)As layer is grown by low-temperature molecular beam epitaxy (MBE) on a GaAs buffer and substrate. Due to the lattice mismatch to the GaAs buffer the (Ga,Mn)As layer is compressively strained and therefore has its magnetic easy axes in the plane perpendicular to the growth direction \cite{shen1997}. After growth of the MBE layers, and without breaking vacuum, the sample is transferred to a UHV evaporation chamber, and 3x0.9 nm of aluminum is deposited on top of the (Ga,Mn)As layer. After deposition, each of the three Al layers is oxidized by keeping it for 8 hours in a 200 mBar oxygen atmosphere. The wafer is then covered by 5 nm Ti and 30 nm Au. The ferromagnetic transition temperature of the (Ga,Mn)As layer is 61 K as determined by SQUID (superconducting quantum interference device).

\begin{figure}
\includegraphics{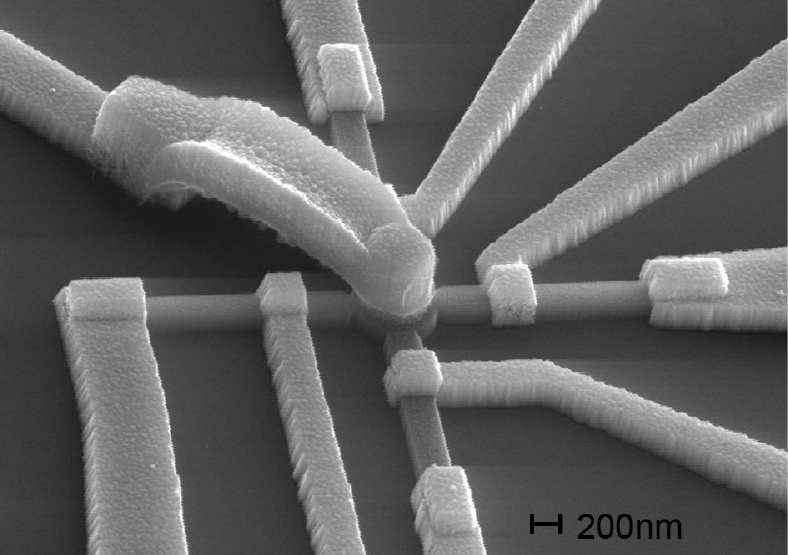}
\caption{\label{fig:SEM} SEM picture of the read-write device. A metallic air bridge out of Gold contacts the central disk. Each of the four nanobars connected to the central disk is contacted by two Ti/Au wires.}
\end{figure}

Figure \ref{fig:SEM} shows the read-write device. It consists of four nanobars which are connected to a circular center region. The structure is defined using electron beam lithography and chemical assisted ion beam etching (CAIBE). The nanobars are 200 nm wide and  2 \textmu m long. After patterning, the Ti/Au and aluminum oxide (Alox) layer are removed from the bars and each nanobar is contacted by Ti/Au contacts using a lift-off technique. The Alox/Ti/Au layer on top of the 650 nm central disk remains on the structure and acts as a read-out tunnel contact. For this purpose, the Au layer on the central disk is contacted by a metallic air-bridge \cite{Borzenko2004}. Small notches are patterned at the transition from the nanobars to the central disk and serve to pin down domain walls.

Thin films of unpatterened compressively strained (Ga,Mn)As exhibit an in-plane biaxial magnetic an\-iso\-tropy at low temperatures. The bars connected to the central disk are aligned with their length parallel to the magnetic easy axes of the bulk. As a result of patterning induced anisotropic strain relaxation \cite{Huempfner2007}, each bar then has a uniaxial magnetic easy axis parallel to its long axis, making the bars appropriate for use of sources for current induced switching of the central disk.

\begin{figure}
\includegraphics{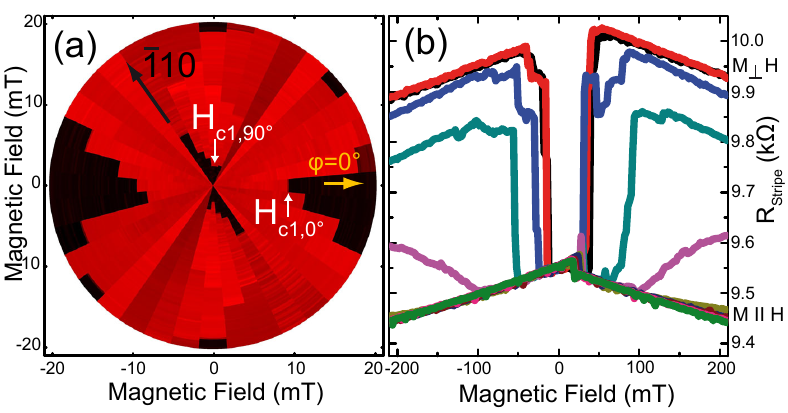}
\caption{\label{fig:AMR} a) Tunneling Anisotropic Magneto Resistance (TAMR) measurements of the central disk summarized in a Polar Plot. b) AMR measurements of one of the nanobars which shows a uniaxial magnetic anisotropy. Magnetic field sweeps for various in-plane angles in steps of $\Delta \varphi=10^{\circ}$ from along the long axis of the nanobar (M$\parallel$H) to perpendicular to it (M$\perp$H).}
\end{figure}

Transport measurements are performed at 4.2 K in a magneto cryostat fitted with three orthogonal Helmholtz coils which can produce a magnetic field of up to 300 mT in any direction. The magnetic anisotropy of each nanobar can be measured in a two terminal configuration. The change in resistance in response to an external magnetic field is due to anisotropic magnetoresistance (AMR) \cite{Jan57,MCGUIRE1975}, which shows a typical $cos^2\vartheta$-dependence where $\vartheta$ denotes the angle between magnetization and current. In (Ga,Mn)As, the resistance for current perpendicular to the magnetization is larger than for current parallel to the magnetization \cite{Baxter2002a}. Fig. \ref{fig:AMR}b shows representative magnetoresistance measure\-ments on one bar for field sweeps along various in-plane directions $\varphi$ referenced to the [100] crystal direction. Such a measurement on a bulk piece of (Ga,Mn)As would show both high and low resistance values at B = 0 mT, as the magnetization would relax to either of the biaxial easy axes depending on the orientation of the magnetic field sweep. In contrast, the resistance of a nanobar at zero magnetic field is always in the low resistance state and independent of the field sweep direction, indicating that, in the absence of an external field, magnetization and current are always parallel to the long axis of the nanobar. The nanobars thus have a uniaxial magnetic an\-isotropy with its magnetic easy axes parallel to the bar.

\begin{figure}
\includegraphics{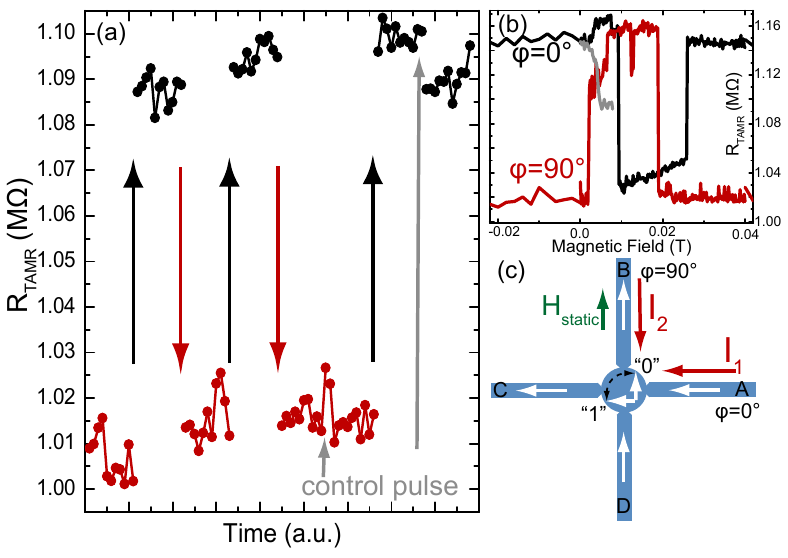}
\caption{\label{fig:MR}a) Switching the magnetization of the central disk due to an electrical current from the $90^{\circ}$ to $180^{\circ}$ direction and back. The device can be written into a "1" ($180^{\circ}$) by applying a current along $180^{\circ}$. For switching back the magnetization in the "0" state one applies the same current in the $90^{\circ}$ direction. Having prepared the "0" state and applying a current in the $90^{\circ}$ direction, the magnetization of the central disk does not change (control pulse) b) full TAMR traces for the central disk ($90^{\circ}$ red $0^{\circ}$ black) and back sweep of the static magnetic field to zero after switching of the magnetization (light gray). c) sketch of the experimental configuration.}
\end{figure}

The relatively large central disk is less influenced by strain relaxation and retains the mainly biaxial anisotropic character of the bulk material. To obtain exact information about the magnetic anisotropies of the central disk we make use of the TAMR (tunneling anisotropic magneto resistance) effect \cite{Gould2004}. The density of states at the Fermi level in (Ga,Mn)As depends on the direction of its magnetization, which leads to a difference in tunnel resistance between the Au and (Ga,Mn)As layers. The TAMR resistance is high for magnetization along [100] ($\varphi=0^{\circ}$) and low for magnetization parallel to [010] crystal direction ($\varphi=90^{\circ}$). Fig. \ref{fig:MR}b (black) shows a TAMR measurement along the $\varphi=0^{\circ}$ direction. The measurement starts with applying -300 mT in the $\varphi=0^{\circ}$ direction and sweeping the magnetic field back to zero. The magnetization at zero field points along $\varphi=180^{\circ}$. Sweeping the field to positive values, the magnetization switches at 9 mT  from $\varphi=180^{\circ}$ to $\varphi=90^{\circ}$ and reverses its direction to $\varphi=0^{\circ}$ at 26 mT. To map the full anisotropy of the central disk, we compile the positive field half of the TAMR-measurements for various directions into a Resistance Polar Plot (RPP) \cite{Pappert2007a}, shown in Fig. \ref{fig:AMR}. Red denotes high and black low resistance. The magnetic field increases with the radius in the RPP and there is high and low (red and black) resistance at zero or small positive fields. For perfectly biaxial material the switching field $H_{c1}$ (where the magnetization first reverses its direction by $\sim90^{\circ}$) along each easy axes would be equivalent. In (Ga,Mn)As, secondary anisotropy contributions cause the two $H_{c1}$ to differ by typically a few percent \cite{Gould2008}. As one can see in Fig. \ref{fig:AMR}a the switching fields $H_{c1}$ in our disk are quite different for $\varphi=0^{\circ}$ and $\varphi=90^{\circ}$ ($H_{c1,0^{\circ}}$= 9.4 mT, $H_{c1,90^{\circ}}$ = 2.0 mT) reflecting a small additional magnetic anisotropy between the two easy axes resulting from strain and patterning.

Having characterized all the individual elements of the structure, we continue with its device operation. To prepare an initial state, we apply a magnetic field $\mu_0 H$ of 300 mT along the $\varphi=120^{\circ}$ direction and sweep it back to 0 mT. As the external field is removed, the device relaxes to a state where (Fig. \ref{fig:MR}c) the magnetization of every nanobar is aligned along its respective long axis. Because it is the biaxial easy axis nearest to the angle of the preparation field, the magnetization of the central disk relaxes to point along $90^{\circ}$ which corresponds to the low resistance state in the TAMR read out (Fig. \ref{fig:MR}b). We define the device to be in its logical "0" state when the magnetization of the central disk points along $90^{\circ}$, and to be "1" when it points along the $180^{\circ}$ direction. To compensate the small magnetic asymmetry of the central disk mentioned above, we apply a static magnetic field of $\mu_0 H=7.8$ mT along $90^{\circ}$. 

Electrical control of the device is then implemented by making use of current induced switching \cite{Yamanouchi2004,Gould2006,Wunderlich2007,Ohno2008}. When a current flows through one of the bars with fixed magnetization, the current carrying holes acquire a polarization, and thus a defined angular momentum. As they pass from the bars into the relatively free disk, the interaction of these carriers with the local Mn moments imparts a torque onto the latter, and for currents above a threshold value, causes the moment in the central region to align to that of the bars from where the current is flowing. 

By choosing the appropriate bars as current source and drain, the magnetization state in our disk can thus be fully controlled. The device is written into a "1"-state (high resistance state) by applying a current with a density of $1\times10^{5} A cm^{-2}$ between contacts A and C of Fig. \ref{fig:MR}c, the magnetization of the central disk switches from the $90^{\circ}$ to the $180^{\circ}$ direction, which results in a high resistance signal for the TAMR read-out contact (Fig. \ref{fig:MR}a). The central disk is switched back by applying a current between contacts B and D. This current-induced switching is clearly detected in the TAMR read-out signal (Fig. \ref{fig:MR}a). The information is written fully electrically and the information storage in the disk is non-volatile. The current density, $1\times10^{5} A cm^{-2}$, is comparable to the current density of ref. \cite{Yamanouchi2004} and one to two orders of magnitude lower than the densities needed in metallic memory elements \cite{pulsetrain}. The TAMR read-out measurement is done with a non-perturbative current of $\sim 1$ nA which does not change the magnetic state of the central disk.

To confirm that the switching of the disk is indeed due to the spin polarization of the current, we prepare again the "0" resistance state. Applying a current along the $90^{\circ}$ direction (contacts B and D) does not change the magnetization of the central disk (see control pulse in Fig. \ref{fig:MR}). Applying the current in the $180^{\circ}$ direction (contacts A and C) switches the central disk to the high resistance state. We have performed similar control experiments for the $180^{\circ}$ direction. The clear outcome of our control experiments is that the switching of the central disk is indeed due to the spin polarization of the current and not due to heating effects, and that we can use current induced switching to control an electrically programmable logic architecture. As a final issue we note that the switching amplitude due to the current is just over 60\% of the full TAMR. This is presumably due to the fact that, when electrically switched, the central disk does not behave as a pure macro spin but allows the formation of domains due to small geometrical imperfections at its edges. To further confirm that the change in resistance has its origin in a switch of the magnetization states, we sweep the static magnetic field back to zero (light gray curve in Fig. \ref{fig:MR}b). This shows that the part of the domain which is not switched was pinned by the applied static magnetic field, and that as soon as we sweep the field back to zero, this part of the disk aligns with the  electrically switched domain. By sweeping back the magnetic field to 300 mT along $180^{\circ}$ the magnetization remains in its position which is clear evidence that the electric current caused a magnetization reorientation of the disk.

As a first step towards a realization of a fully a programmable logic device, we described recently in ref. \cite{Pappert2007b} an ultra-compact (Ga,Mn)As based memory cell. In that work, we made use of lithographically engineered strain relaxation \cite{Huempfner2007} to produce a structure comprised of two nanobars with mutually orthogonal uniaxial easy axes, connected by a narrow constriction. Measurements showed that the resistance of the constriction depends on the relative orientation of the magnetization in the two bars. While very small, the functionality of the memory cell was dictated by its lithographic layout. We will now describe how a universal gate can be fabricated in (Ga,Mn)As by adding bulk biaxial anisotropy and writing of the information by electrical means to the initial concept. 

\begin{figure}
\includegraphics{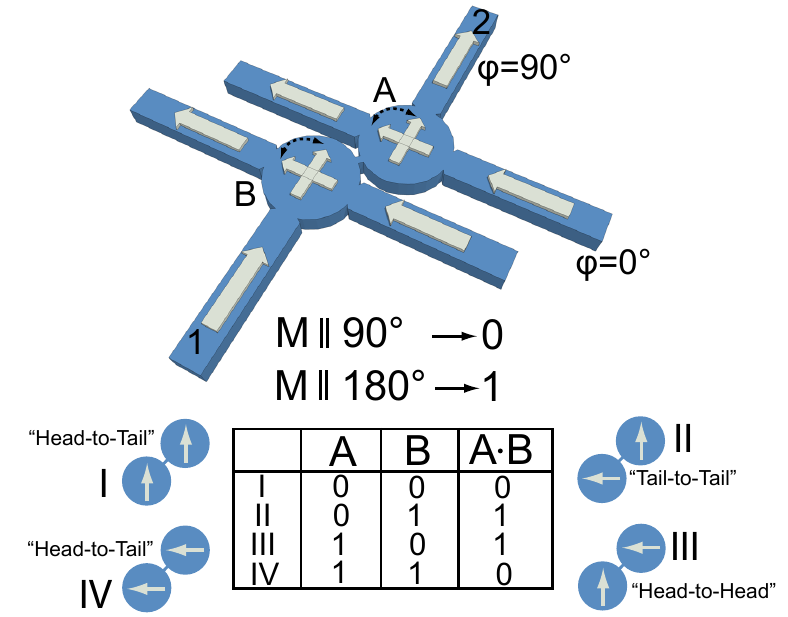}
\caption{\label{fig:logicelement}Proposed programmable logic element. The two central disks serve as two memory bits as well as two input bits of the logic device. The constriction between the two central disks is the key to reading-out the result. The resistance from contact 1 to 2 will be dominated by the constriction which is dependent on the magnetization configurations of the input bits. In this configuration the input bits have four possible magnetization states: twice "Head-to-Tail", "Head-to-Head", "Tail-to-Tail". Truth table as it could be programmed as an exclusive OR (A XOR B) gate.}
\end{figure}

Extending our read-write device from Fig. \ref{fig:SEM} to two central disks and connecting the two disks with a small constriction creates a fully electrically programmable logic and storage device. The blue shape in Fig. \ref{fig:logicelement} depicts the design of such a logic device. Two central disks act as non-volatile storage units and at the same time represent the input of the two bit logic operation. The 'bit value' is represented by the in-plane magnetization direction of each disk.  The element is initialized by applying a field in a specific direction and sweeping it back to zero. For example, for initializing along $120^{\circ}$, the resulting magnetic orientation of the bars is as given by white arrows in Fig. \ref{fig:logicelement} and the initial magnetization of both central disks points in the $90^{\circ}$ direction. During device operation, the configuration of each central disk can be changed into two possible magnetization directions pointing either in the $180^{\circ}$ or $90^{\circ}$ direction. A current flowing from a given bar into the disk switches the magnetization of the disk parallel to the magnetization of that bar. The constriction between the two central disks is the key to reading-out the result. As long as the connection is sufficiently narrow, the resistance between contacts 1 and 2 in Fig. \ref{fig:logicelement} will be dominated by this constriction. The resistance of this constriction depends on the relative magnetization states of the central disks \cite{Pappert2007b}. If the magnetization of both disks point either towards or away from the constriction we call the configuration Head-to-Head or Tail-to-Tail, respectively as depicted in the insets of Fig. \ref{fig:logicelement}. The magnetic field lines caused by the magnetization of the disks are perpendicular to the current direction in the constriction. On the other hand, if the magnetization of the disks is in series (magnetization of one disk pointing toward the constriction and the other disk pointing away from the constriction, Head-to-Tail) the magnetic field lines in the constriction are parallel to the current direction. Because of an effect akin to anisotropic magnetoresistance (AMR), and associated with a magnetization dependence of the impurity wave functions in (Ga,Mn)As  \cite{Pappert2007b,Schmidt2007}, the resistance of the constriction depends on the angle between the field lines and the current through the constriction allowing to determine the relative magnetization states of the two bits. The output of the logic operation is defined as "1", if the magnetic configuration of the disks is Head-to-Head or Tail-to-Tail and is "0", if the states are in Head-to-Tail configuration. 

For e.g. an exclusive OR (A XOR B) logic element we define the magnetization direction of the disks pointing in $90^{\circ}$ as "1" and pointing along $180^{\circ}$ as "0". For the initial configuration both disks are in the "1" state and the magnetic configuration of the output is Head-to-Tail and therefore "0". A switching current through disk "A" along $180^{\circ}$ switches the magnetization of disk "A" along $180^{\circ}$ changing the "A" input to "0". The relative magnetization is now  Head-to-Head, as symbolized by configuration III in Fig. \ref{fig:logicelement}  and the XOR logic operation yields "1" as an output (III: 1 XOR 0 = 1). If disk B is also switched to the $180^{\circ}$ direction the device has again Head-to-Tail configuration and therefore yields "0" as output. The complete truth table is given in Fig. \ref{fig:logicelement}.

The present results for electrically writing information into the read-write device, combined with the constriction read-out results of ref. \cite{Pappert2007b}, provide all essential elements for the realization of our programmable logic element. In closing we stress that the functionality of the programmable logic scheme presented here can be straightforwardly extended to produce multi-purpose functional elements \cite{Patent07}, where the given geometry can be used as various different computational elements depending on the number of input bits and the chosen electrical addressing. Such a paradigm has technological advantages as it allows for the generation of entire computational circuits consisting of multiple identical elements, which can thus be easily, rapidly and cheaply produced by parallel lithography.

\vspace{0.1cm}
\begin{acknowledgments}
The authors thank T. Borzenko and V. Hock for help in sample fabrication and M. R\"uth and T. Naydenova for useful discussions. They also acknowledge financial support from the German DFG (Schm1532/5-1, Br1960/4-1) and the EC (NANOSPIN FP6- IST-015728, SemiSpinNet FP7-215368).
\end{acknowledgments}


\begin{thebibliography}{10}
\bibitem{pulsetrain}
\textnormal{We observe that the switching is better controlled when the current
  is delivered as a series of 10 pulses, each 40 \textmu s long, during a 5 s
  time frame.}
\end{thebibliography}
\end{document}